\documentclass[fleqn,10pt]{wlscirep} 
\usepackage{amsmath,bm}

\title{Hypersonic  Bose-Einstein Condensates in Accelerator Rings}

\author[1,2]{Saurabh Pandey}
\author[1,3]{Hector Mas}
\author[1,2]{Giannis Drougakis}
\author[1]{Premjith Thekkeppatt}
\author[1]{Vasiliki Bolpasi}
\author[1]{Georgios Vasilakis}
\author[1]{Konstantinos Poulios\footnote{Current Address: School of Physics \& Astronomy, University of Nottingham, University Park, Nottingham NG7 2RD, UK}}
\author[1,$\dag$]{Wolf von Klitzing}
\affil[1]{Institute of Electronic Structure and Laser, Foundation for 
Research and Technology-Hellas, Heraklion 70013, Greece}
\affil[2]{Department of Materials Science and Technology, University of Crete, Heraklion 70013, Greece}
\affil[3]{Department of Physics, University of Crete, Heraklion 70013, Greece}

\affil[$\dag$]{Email: wvk@iesl.forth.gr}

\begin{abstract}
\textbf{
Some of the most sensitive and precise measurements to date are based on matterwave interferometry with freely falling atomic clouds.
Examples include high-precision measurements of inertia \cite{Gustavson1997PRL}, gravity \cite{Rosi2014N} and rotation \cite{Dutta2016PRL}. 
In order to achieve these very high sensitivities, the interrogation  time has to be very long and consequently the experimental apparatus has to be very tall, in some cases reaching ten or even one hundred meters \cite{Kovachy2015N, Zoest2010S}.
Cancelling gravitational acceleration, e.g. in atomtronic circuits \cite{Amico2017NJOP, Dumke2016NJP} and matterwave guides \cite{Wang2005PRL}, will result in compact devices having much extended interrogation times and thus much increased sensitivity both for fundamental and practical measurements.
In this letter, we demonstrate extremely smooth and controllable matterwave guides by transporting Bose-Einstein condensates (BEC) over macroscopic distances: We use a novel neutral-atom accelerator ring to bring BECs to very high speeds (16x their velocity of sound) and transport them in a magnetic matterwave guide for 15 cm whilst fully preserving their internal coherence.
The high angular momentum of more than 40000\,$\hbar$ per atom gives access to the higher Landau levels of quantum Hall states.
The hypersonic velocities combined with our ability to control the potentials with pico-Kelvin precision open new perspectives in the study of superfluidity and give rise to new regimes of tunnelling and transport \cite{Brantut2012Science, Krinner2017JoP, Krinner2015Nature}.
Coherent matterwave guides are expected to enable interaction times of several seconds in highly compact devices. These developments will result in portable guided-atom interferometers for applications such as inertial navigation and gravity mapping.
}
  \end{abstract}
\begin{document}

\flushbottom
\maketitle

\thispagestyle{empty}

%=================================================================================
%=========================== MAIN PART ===========================================
%=================================================================================
Ring-shaped atom circuits are excellent candidates for guided matterwave Sagnac interferometry \cite{LENEF1997PRL}, Josephson-oscillations of angular momentum \cite{Lesanovsky2007PRLb}, and atomtronic applications such as the quantized conductance through a constriction 
\cite{Brantut2012Science, Krinner2017JoP, Eckel2014N,   Krinner2015Nature}.
In the past decades, a number of ring traps have been implemented mainly based on magnetic \cite{Murch2006PRL, Navez2016NJOP} and optical dipole trapping \cite{Henderson2009NJOP}.
In small traps the deleterious effects of the corrugations can be avoided by operating at velocities well below the critical velocity of the superfluid  BECs. 
This has allowed ring shaped traps to be used in fundamental studies, e.g.\,on the correspondence between superfluidity and Bose-Einstein condensation \cite{RYU2007PRL} and on the hysteresis of flux in a ring atomtronic circuit \cite{Eckel2014N,RYU2007PRL}. 
Atom-chip based magnetic waveguides, on the other hand, have been a very successful platform for many cold atom experiments requiring non-trivial geometries \cite{Wang2005PRL} and have  been used to study the propagation of BECs in matterwave guides 
\cite{Leanhardt2002PRL}.
One of the main problems of chip based waveguides and complex optical potentials is that small modulations in the confining potential are almost unavoidable \cite{Trebbia2007PRL}.
Methods have been developed to reduce the effect of this roughness, e.g.~by modulating the currents in the microchip traps \cite{Trebbia2007PRL} or by using optimal control theory  \cite{CHEN2011PRA,Guery-Odelin2014PRA}.
Most previous experiments were performed in a regime, where the effect of the roughness of guiding potential is suppressed by the superfluidity of the BECs. 
Atom interferometers, however, need to operate at high speeds and low atom densities, where the energy shifts associated with the superfluid properties are vanishingly small.  
Therefore, the presence of any corrugation or roughness of the guiding potential of atom interferometers leads to a coupling of the forward momentum to transversely excited states---thus scrambling the phase of the interferometer and severely limiting the  distances over which the atoms can be guided coherently 
\cite{Leanhardt2002PRL}.  

\begin{figure}[t]
\begin{center}
  \includegraphics[width=0.8 \textwidth,angle=0]
  {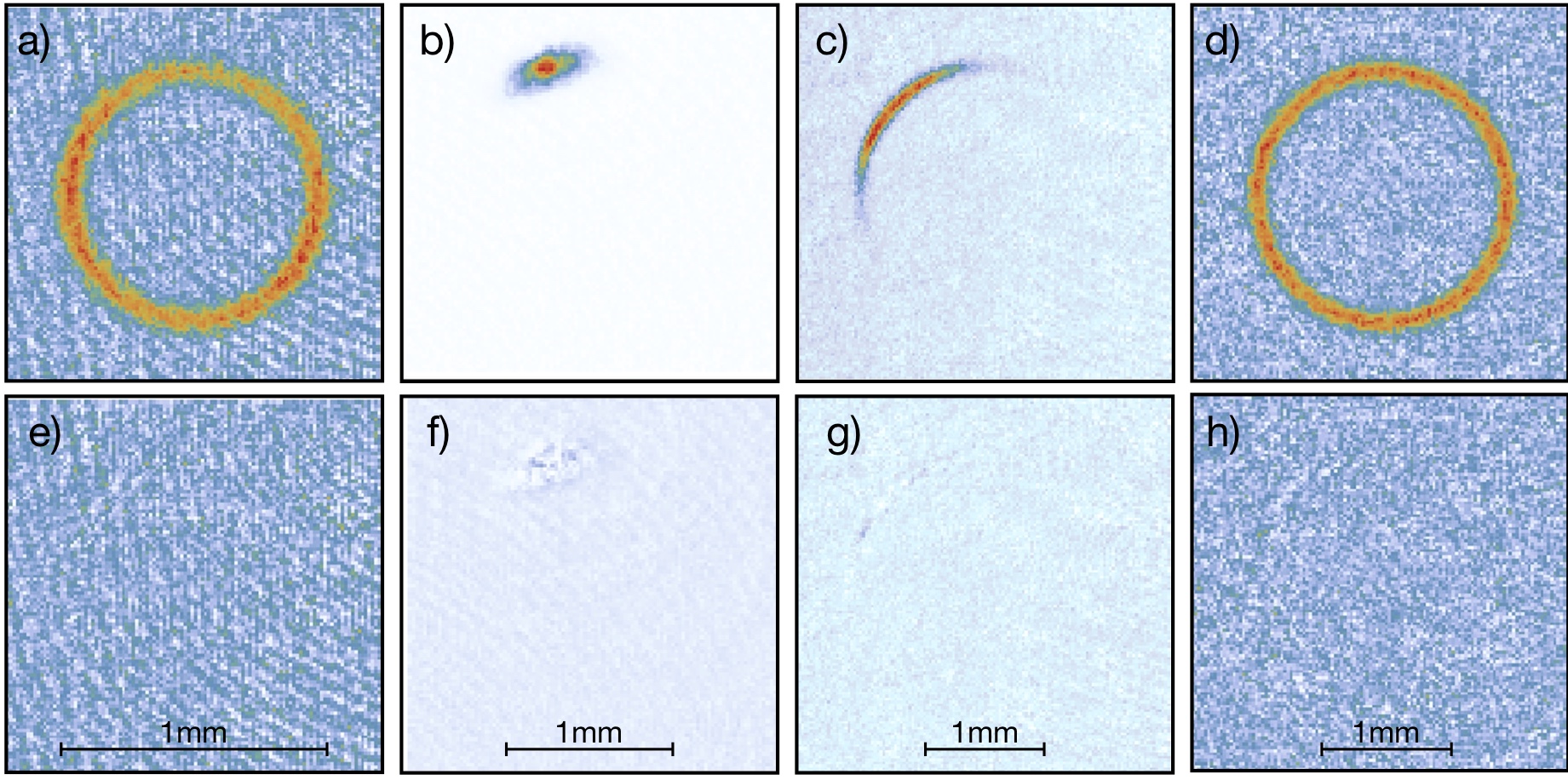}
 \caption{\textbf{Absorption images of ultra-cold thermal clouds and BECs in ring-shaped matterwave guides}. The upper row of images show the column density of the atoms in the ring-shaped waveguides. In a) the atoms are at rest, while b), c), and d) show atoms travelling at maximum velocity. 
 b) shows the atoms  moving in the acceleration potential along the waveguide. c) shows the atoms just after being released into the waveguide, and d) shows the atoms after travelling freely for 2\,s in the waveguide.
 \\The lower row shows the experimental background densities, calculated by subtracting from upper images  a smooth theoretical model of the atomic densities. 
 The complete absence of any signature of the ring in e) and h) clearly demonstrates that there is no detectable roughness in the atomic distribution and this in wave-guide potential. 
 From the fits to a) and d) we deduce a maximum effective modulation of the potential of 250\,nK and 189\,pK respectively.}
\label{fig:flatrings}
\end{center}
\end{figure}

In this work, we report the first experimental realisation of matterwave guiding of BECs over macroscopically large distances whilst completely preserving its internal coherence.
In our ring-shaped waveguides, BECs  travel at hypersonic speeds of 28\,mm/s for as much as 148\,mm without any appreciable additional heating or reduction in lifetime as compared to the static case.
The peak-to-peak roughness of our waveguide is smaller than our measurement limit of 189\,pK, which corresponds to a maximum difference in gravitational potential of less than 2\,nm over the whole ring (radius $R=443\,\mu$m).
The traps and waveguides presented here are based on time-averaged adiabatic potentials (TAAPs), which rely for the definition of the shape of the waveguide on a simple DC quadrupole field plus the polarisation and amplitudes of homogeneous fields oscillating at audio and radio frequencies. 
The field-generating coils are large and distant when compared to the atomic ensemble.
This limits the highest spatial frequency of the trapping potential in the azimuthal direction to $4\pi R$, i.e.~a maximum of two minima per turn.
Any harmful corrugation or roughness due to imperfections in the wires falls off exponentially with the distance from the coils ($\sim$50\,mm) divided by the relevant length scale ($\ll 1$\,mm), resulting in perfectly smooth TAAP waveguides\footnote{The unavoidable  imperfections of the field generating coils result in a modulation of the magnetic field close to its surface.
The resulting corrugation in the trapping potential at a spatial frequency $k=2\pi/\lambda$  dies off as $\exp(-k z)/\sqrt{k z}$ over the distance z~
\cite{Jones2004JOPB}.
We take as the minimum spatial frequency of interest the inverse of the diameter of the  ring shaped waveguide of about than 1\,mm$^{-1}$.
With coils being at a distance of 50\,mm from the atoms, this implies a reduction in coil-induced defects by a factor of $3\times 10^{-23}$.
}.
This has to be compared to atom-chip based traps which use the shape of close-by wires to define the trapping potential \cite{Folman2000PRL}
and optical dipole potentials \cite{Henderson2009NJOP} 
using the spatial distribution of the light.
Both can create much more complex structures along the waveguide and thus have small but unavoidable corrugations of the wave-guiding potential.
Figure\,\ref{fig:flatrings} shows absorption images of atoms in a ring-shaped TAAP waveguide with Fig.\,\ref{fig:flatrings}a showing static atoms, and Fig.\,\ref{fig:flatrings}b, c and d showing atoms guided at high speeds.
The images in lower panels show the error of fits of a smooth potential to the images directly above them. 
Note that there is no  trace visible of the ring in the error,  thus demonstrating the smoothness of the atom distribution and thus of the waveguide itself.

Our TAAP traps and waveguides are generated by  combining AC and DC magnetic fields at three different time-scales: DC, audio-frequency, and RF-frequency  \cite{LESANOVSKY2007PRL,
SHERLOCK2011PRA,Navez2016NJOP}. 
The RF-fields $(\bm{B}_\text{\textbf{rf}}(t))$ dress the  states of the atoms in the DC magnetic quadrupole field $(\bm{B}_\text{\textbf{dc}})$  yielding adiabatic states, which are time-averaged by a spatially homogeneous  magnetic field $(\bm{B}_\text{\textbf{m}}(t))$ oscillating at audio-frequencies  $(\omega_\text{m})$.
The frequency of the time-averaging field  is chosen such that the magnetic spin of the atoms can follow adiabatically but the centre of mass of the atoms remains virtually unchanged ($\omega_\text{m}/2\pi=5$\,kHz) \cite{LESANOVSKY2007PRL}.
Starting from a DC magnetic quadrupole trap and a vertically polarised radio-frequency field, we can produce 
a ring-shaped waveguide by applying a vertical audio-frequency field.
If we tilt the audio field from its vertical axis, then the ring tilts as well and gravity creates an azimuthal trapping potential along the ring.
The full time-dependent magnetic field is 
\begin{equation}
\bm{B}\, = \,\, \alpha \left( {\begin{array}{{c}}
  x \\ 
  y \\ 
  { - 2z} 
\end{array}} \right) 
+{B_{\text{m}}}\sin {\omega _{\text{m}}}t
\left( {\begin{array}{{c}}
  { \delta \cos \phi_0} \\ 
  { \delta \sin \phi_0} \\ 
  1 
\end{array}} 
\right)
+{B_{{\text{rf}}}}\sin {\omega _{{\text{rf}}}}\,t \left({\begin{array}{{c}}
  0 \\ 
  0 \\ 
  1 
\end{array}} \right),
\label{eq:B-Full}
\end{equation}
where the first term of the sum stands for the DC quadrupole $(\bm{B}_\text{\textbf{dc}})$ field of gradient $\alpha$.
The second term represents the audio-frequency modulation field $(\bm{B}_\text{\textbf{m}}(t))$ which is mainly in the vertical (z-) direction but can be tilted by a small angle $\delta$ in the direction $\phi_0$.
The final term of the sum is the RF-field $(\bm{B}_\text{\textbf{rf}}(t))$ with a linear polarisation in the vertical direction. 
Near the resonance  $\hbar\omega_{\text{rf}}= |g_{\text{F}}| \mu_{B} B$ the rf field $\bm{B}_\text{\textbf{rf}}(t)$ dresses the atoms in the magnetic field, which turns the  quadrupole into a shell-like trap \cite{Zobay2001PRL}.
If the modulation frequency $(\omega_\text{m})$ of the homogeneous field $\bm{B}_\text{\textbf{m}}(t)$ is small compared to the Larmor frequency $(\Omega_{\text{L}}=\mu_{B}|g_{F} \bm{B}|)$ but high compared to the eventual radial trapping frequency  $(\omega_{\text{z},\text{r}})$, then the atoms are trapped in the time-average of the adiabatic potential, which results in a ring-shaped matter-waveguide of radius $R \sim \hbar \omega_{\text{rf}}/ |g_{\text{F}}| \mu_{\text{B}} \alpha$ from the zero field point \cite{LESANOVSKY2007PRL}. 
A detailed description of the TAAP potentials is described elsewhere \cite{LESANOVSKY2007PRL,Navez2016NJOP}. 
Near the core of the waveguide, the TAAP ring potential can be described in cylindrical coordinates $(r,\phi,z)$  as \cite{LESANOVSKY2007PRL, Navez2016NJOP}:
\begin{equation}
{V_{\text{r}}(r, \phi, z)} = \hbar{\Omega_{\text{rf}}} + \frac{1}{2}m{\omega_{\text{r}}^{2}}(r-R){^2} +\frac{1}{2}m{\omega_{\text{z}}^{2}}z^2-\frac{\delta }{2}m g R \cos\left({ \phi-\phi_0} \right),
\label{eq:SimpleRingPotential}
\end{equation}
 where ${\Omega_{\text{rf}}}$ is the Rabi frequency associated with the RF-field, ${\omega_{\text{r}}}$ and ${\omega_{\text{z}}}$, 
are the radial and the axial trapping frequencies, respectively.
For $\delta=0$, Eq.~\ref{eq:SimpleRingPotential} represents a circular waveguide. 
Fig.\,\ref{fig:flatrings} shows an ultra-cold atomic clouds in such ring-shaped potentials.
The radial and axial confinement frequencies are ${\omega_{\text{r}}} = {\omega_0} (1+{\beta_{\text{m}}^{2}}){^{-1/4}}$ and ${\omega_{\text{z}}} = 2 {\omega_0} [1-(1+{\beta_{\text{m}}^{2}}){^{-1/2}}]{^{1/2}}$ respectively, where $\beta_{\text{m}} = |g_{\text{F}}| \mu_{\text{B}} B_{\text{m}}/\hbar {\omega_{\text{rf}}}$ is the index of modulation of the time-averaging field, and ${\omega_0} = (|m_F g_{\text{F}}| {\mu_{\text{B}}}\alpha) (m \hbar {\Omega_{\text{rf}}}){^{-1/2}}$ is the radial trapping frequency of an adiabatic shell potential in the absence of modulation \cite{LESANOVSKY2007PRL,Zobay2001PRL}.

One of the most interesting aspects of a circular waveguide is its ability to guide atoms with extreme precision at very high angular momentum. Fig.\,\ref{fig:flatrings}d, for example, shows atoms travelling at an angular momentum of $L=m  R^2\dot\phi =17000\,\hbar$ per atom. 
This is an ideal starting point, e.g.~for the excitation of quantum Hall states and well-defined higher-lying Landau levels.
In order to accelerate the atoms one needs to create an azimuthal potential along the ring. 
For this we apply a small horizontal modulation field, which tilts $\bm{B}_\text{\textbf{m}}(t)$ and thus the ring 
 in the direction of the  horizontal modulation field $(\phi_0)$ by an angle $\delta /2$ with respect to horizontal  (see Eq.~\ref{eq:B-Full} with $\delta \ne 0$).
The gravitational potential of the atoms then creates a trap in the azimuthal direction at $\phi_0$. 
The azimuthal trapping frequency is then simply the one of a tilted pendulum ${\omega_{\phi}} = \sqrt{\delta g/2R}$. 
By adjusting the amplitudes  of the modulation fields in the two horizontal Helmholtz coil pairs we can freely move the position $\phi_0$ of the minimum around the ring.
The basic idea of our accelerator ring is to load a BEC into a static, tilted ring  and then modulate the amplitudes of the modulation field in the x- and y- directions such that the minimum of the trap accelerates along the ring and then transports the BEC at constant angular velocity over large distances. 

A sudden acceleration, however,  would excite centre of mass oscillations of the ultra-cold cloud.
Adiabatic acceleration on the other hand would take prohibitively long. 
An elegant solution is provided by optimal control theory and its  so-called  \emph{`bang-bang'} scheme\cite{CHEN2011PRA},
which compensates the force due to a constant acceleration by an opposite force due to an offset in the position of the atomic cloud relative to the center of the moving harmonic trap.
We do this by jumping the position of the trap instantaneously forward by $ \Delta \phi = + \ddot\phi/\omega{^2}$ exactly at the moment when we start the acceleration.
In the accelerated frame, the atoms then stay  exactly at the bottom of the effective trapping potential.
Once the target velocity is reached one abruptly stops the acceleration $(\ddot \phi=0)$ and jumps the phase back by $-\Delta\phi$ thus placing the atomic cloud at the bottom of the trap moving at constant velocity.
\footnote{The uniform acceleration is a uniform force on the atoms, which in turn corresponds to a shift of the parabolic azimuthal trapping potential. We correct this shift by jumping the trapping potential in the opposite direction. However, in practice the trapping frequency is not entirely independent of the angular velocity, which necessitates fine-adjustment of the size of the phase jump.}. The same method can be used to decelerate the atoms.
\begin{figure}[!ht]
\begin{center}
  \includegraphics[width=0.6 \textwidth,angle=0]{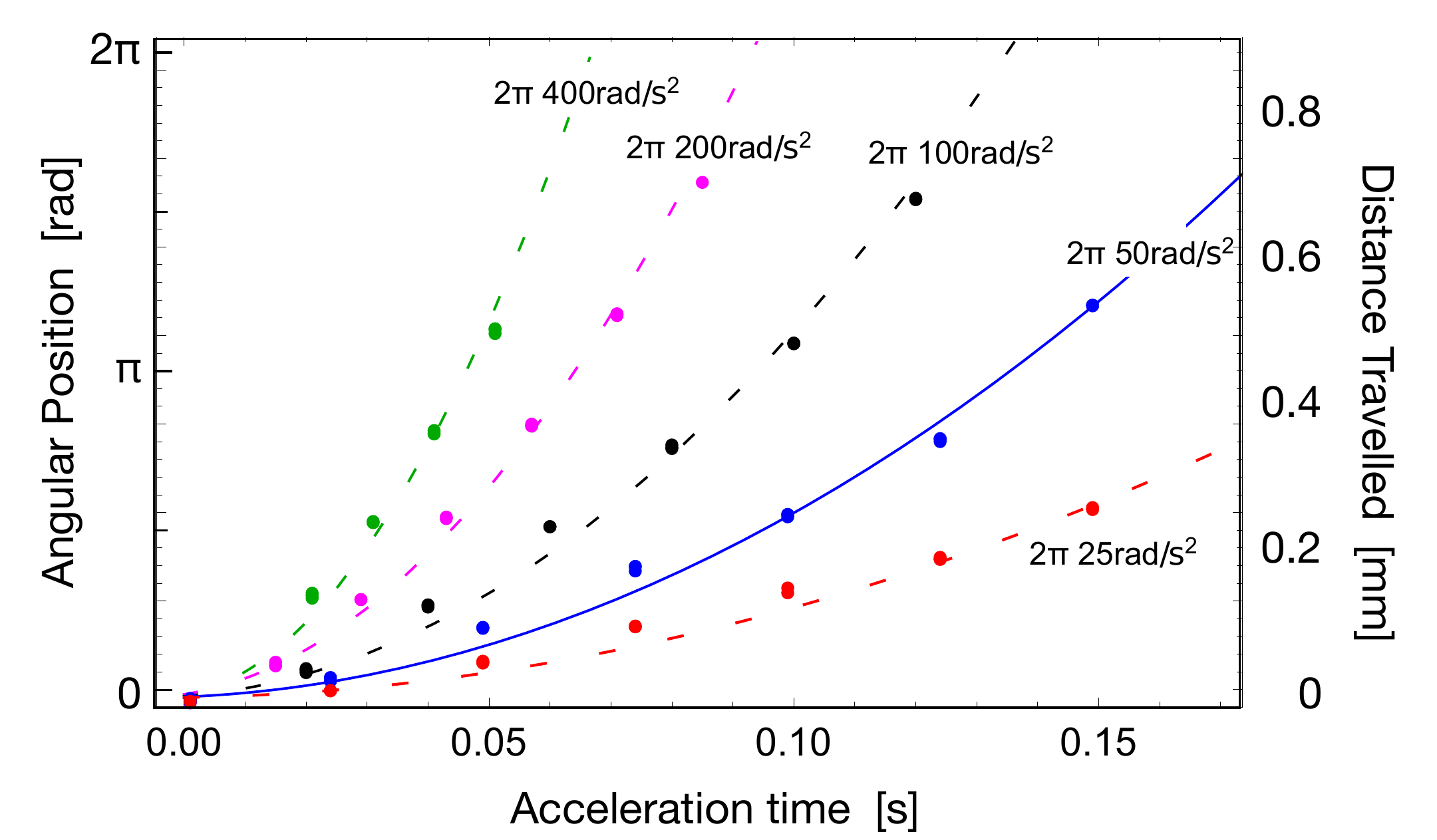}
 \caption{
 \textbf{Accelerating BECs}: BEC angular position vs time in a tilted ring trap for five different angular accelerations. $x, y$ modulation amplitude is 0.55 G and 1.4 G along the $z$ axis. The lines are theoretical prediction of the position of the clouds $(\phi=\frac{1}{2}\ddot\phi t^2)$. 
 The error in the data is smaller than the point size. 
 The uncertainty in the fitted angular position and distance travelled result from an uncertainty in the position of the center of the ring and the magnification of the imaging system  (27\,mrad or 2\% of distance).
The continuation of the solid line and its corresponding data points is plotted in Fig.\,\ref{fig:transport}.
}
\label{fig:accelerator}
\end{center}
\end{figure} 
Figure \ref{fig:accelerator} shows the measured position of BECs (dots)  along with its theoretical prediction (lines) during the acceleration in a TAAP ring, which ranged from $2\pi\,25$\,rad/s$^2$ to $2\pi\,400$\,rad/s$^2$ reaching angular velocities of up to $2\pi\,20$\,rad/s corresponding to an angular momentum of 44,600\,$\hbar$.
The deviation from the theoretical curves result from small changes in the trapping frequency during the acceleration.  
The centrifugal force associated with the rapid rotation in the ring forces the atoms outward from  a radius of $436(2)\mu$m in the static trap to $443.4(4) \mu$m  at an angular speed of $2\pi 10$\,rad/s.
Care has to be taken that the  change in trapping frequency  does not cause parametric heating of the sample. 
Here, for the acceleration of  $2\pi\,50$\,rad/$s^2$ to a final angular velocity of  $\dot \phi=2\pi\,10\,$Hz the azimuthal trapping  frequency decreases within 0.2\,s from $\omega_\phi=2\pi\,9.17(3)$\,Hz to its final value of $2\pi\,$7.76(1)\,Hz. 
Therefore, $\dot{\omega}_\phi / \omega_\phi^{2}=0.02\ll 1 $, which means that this change in trapping frequency is fully adiabatic even in the azimuthal direction \cite{LESANOVSKY2007PRL}.
Nevertheless, the phase jumps have to be optimised to take this change of trapping frequency into account.
By making small adjustments of the phase jumps both at the beginning and the end of the acceleration, we can suppress any oscillation of the condensate in the final trap.

\begin{figure}[!ht]
\begin{center}
  \includegraphics[width=0.97 \textwidth,angle=0]
  {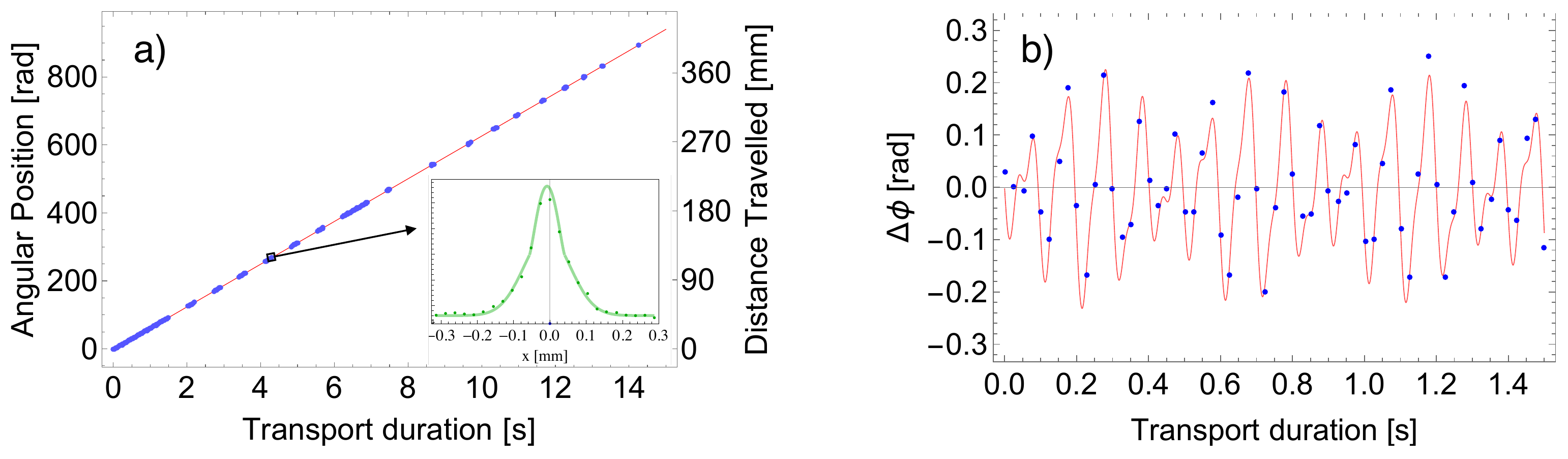}
 \caption{\textbf{Long distance transport in the accelerator ring.} a) The angular position of the condensate and thermal cloud during  14.3\,s of transport in the matterwave guide. The red line depicts the programmed trajectory of $2\pi\,10$\,rad/s. The inset shows the bi-modal distribution of the Bose-Einstein condensate after 4.1\,s of transport and 24\,ms time-of-flight expansion. b) is the condensate angular position relative to the programmed trajectory at $2\pi\,10\,\text{rad/s}$. The oscillations are partially due to a small azimuthal modulation of the trapping potential and partially due to a small center of mass oscillation of the cloud relative to the moving trap.
}
\label{fig:transport}
\end{center}
\end{figure}

In the experiments described here, we start with BECs of $3\times 10^5$ atoms at 32\,nK in a static trap in the ring, accelerate them at $2\pi\,50$\,rad/${s^2}$ for 200\,ms,
and continue the transport at a constant angular speed of $\dot \phi_\text{f}=2\pi\,10\,\text{rad/s}$ for up to 14.3\,s 
or 39.8\,cm (see Fig.\,\ref{fig:transport}). 
The phase jumps were  $\Delta \phi=\pm0.5$\,mrad, respectively. 
The inset of Fig.\,\ref{fig:transport}a shows a bi-modal fit to the atomic density after 41 round trips corresponding to a total transport distance of 11.4\,cm. 
Fig.\,\ref{fig:transport}b shows the deviation of the position of the atoms from the programmed trajectory at a constant speed of $2\pi\,10\,\text{rad/s}$.
The angular position of the atoms was fitted to the sum of three sine waves: Two at the fixed rotation frequency $2\pi\,10\,\text{rad/s}$ and its first harmonic $2\pi\,20\,\text{rad/s}$, plus one decaying sine wave of arbitrary frequency. 
The fitted amplitudes were 140(10)\,mrad, 40(10)\,mrad, and 70(10)\,mrad, respectively. 
The third sine wave corresponds to an azimuthal centre of mass motion of the atomic cloud in the moving trap at 7.76(1)\,Hz, decaying with a 1/e time constant of 5.3 seconds.
This very precise measurement of the centre of mass oscillation (see Fig.\,\ref{fig:transport}b) can be used to eliminate any final oscillation in the moving trap by fine-tuning the phase jumps at the beginning and end of the acceleration.
The extremely fine control of the velocity of the atoms in the TAAP accelerator ring for neutral atoms could be used for highly-controlled atomic collision experiments.
The 10\,Hz and 20\,Hz modulations originate from the non-perfect flatness of the ring.
A fit to a static thermal cloud in Fig.\,\ref{fig:flatrings}a)  reveals a slight modulation of the azimuthal potential at angular frequencies of $2\pi$ (tilt of the ring) and $4\pi$ (polarisation induced warping).  
This  reappears during the transport as a micro-motion \cite{
MULLER2000PRL 
} in the moving trap at exactly the final angular velocity and its harmonic.

If Bose-Einstein Condensates move slower than their speed of sound, then superfluidity  allows  them to flow  around obstacles without dissipation or heating. 
This is the regime where most guided condensates have been operated so far \cite{KUMAR2016NJP,Eckel2014N,RYU2007PRL,Lesanovsky2007PRLb,Krinner2015Nature,Leanhardt2002PRL}.
At larger velocities though, any corrugation of the guiding potential couples the forward motion of the atoms to their transverse degrees of freedom, leading to a distortion of the BEC, heating, or atom loss and eventually to the destruction of the BEC \cite{Murch2006PRL,Leanhardt2002PRL}.
Therefore, one can probe the roughness of a guiding potential by propagating BECs at  high velocities.
For our TAAP waveguides, we do not observe any change in lifetimes of the thermal clouds  (3.3\,s)  and BECs (5.3\,s), when 
comparing the atomic clouds in the moving and static traps \footnote{The thermal lifetime corresponds to the point where the thermal atom number has dropped to 1/e of its original value, whereas the BEC lifetime is the time it takes for the BEC to vanish completely.}.
The measured heating rate of $3\pm1\,$nK/s is the same in the moving and static cases, giving an upper limit to the additional heating rate induced by the guiding of smaller than 1\,nK/s or 32\,pK/mm for condensates travelling at a speed of 28\,mm/s.
A BEC lifetime of 5.3\,s is particularly impressive since the atoms travel in the waveguide at a hypersonic speed of 16 times the Landau critical velocity in the BEC \footnote{At its final angular speed of $2\pi\times 10$\,rad/s the BEC is travelling at 2.8\,cm/s.
The Landau critical velocity of a superfluid of $3\times 10^5$ atoms in a trap of $46\times85\times7.8\,$Hz$^3$  is about 0.17\,cm/s, which means that the atoms are travelling at Mach 16. See also the Methods section.}.
At such high velocities any roughness or corrugation of the guide would couple the longitudinal velocity to transverse excitations and thus rapidly destroy the coherence of the condensate.
The absence of any measurable heating thus is only possible because the TAAP matterwave guides are perfectly smooth.

To our best knowledge, this is the first demonstration of lossless hypersonic transport of condensates in a matterwave guide. 
Such smooth guides form an ideal base for guided matterwave interferometry. The circular waveguide lends itself naturally to Sagnac interferometry, where the atoms take one or multiple round trips in opposite directions. 
Atoms propagating through two opposite halves of the ring form a Michelson interferometer, which---due to the potentially very long interaction time---can be made superbly sensitive to gravitational acceleration and even gravitational waves.
The ability to create BECs at extremely high and well controllable angular momentum per atom is the ideal starting position for highly-correlated angular momentum states such as Laughlin or quantum Hall states \cite{Roncaglia2011SR}.
The transfer to the necessary harmonic trap can be achieved by simply ramping down the amplitude of the vertical time-averaging field and thus adiabatically transferring the atoms from the ring trap to the harmonic bottom of a shell type trap.

To explore this further, we  study the free propagation of a BEC in the waveguide eventually filling the full ring.
We start by characterising the flatness of the ring potential  ($\delta$=0) using a static cloud. 
We load $10^6$ thermal atoms at 430\,nK in a static ring trap of radius $470\,\mu$m and allow them to thermalise  for 6.5\,s.
By imaging the atoms, we can use the Maxwell-Boltzmann distributions $(f\propto Exp[-U(\phi,\rho)/kT)])$ to determine the potential energy landscape.
For sufficiently long expansion times, i.e.\;$ t_\text{exp} \gg \omega_{\text{r}}^{-1}$, the radial atomic distribution yields the temperature. 
For sufficiently short expansion times, i.e.\;$ t_\text{exp}\ll \omega_\phi^{-1}$, the azimuthal distribution of the atoms provides the azimuthal shape of the trapping potential.
For $\omega_{\text{r}}^{-1}\ll t_\text{exp}\ll \omega_\phi^{-1}$ both can be achieved in a single absorption image like those in  Fig.\,\ref{fig:flatrings}.
The azimuthal dependence the theoretical description of TAAP rings for arbitrary modulation and RF-polarisation\cite{Navez2016NJOP} can be rewritten using only the first two cylindrical harmonics.  
The azimuthal potential can therefore be parameterized as
\begin{equation}
    V_\phi=V_{\phi c}\left[1 + {h_1} \cos(\phi + {\phi_1}) + {h_2} \cos(2 \phi + {\phi_2})\right],\label{equ:azimuthalFit}
\end{equation}
 with arbitrary amplitudes $h_1$ and $h_2$ and angular positions $\phi_1$ and $\phi_2$ \cite{Navez2016NJOP}. 
We fit our experimental images to a Maxwell-Boltzmann distribution for $V_\text{r}+V_\phi$ (see Methods section).
The fact that the fit-residual  in Fig.\ref{fig:flatrings}e contains  no trace of the original ring in  Fig.\ref{fig:flatrings}a shows that the highest spatio-angular frequency present in the azimuthal potential is indeed $2\phi$, further supporting the extreme smoothness of the TAAP potentials.
The fit yields $2\times 10^5$ atoms at 502\,nK and the modulation amplitudes of $h_1=0.11$ and $h_2=0.20$ at $\phi_1$=-118$^{\circ}$ and $\phi_2$=+115$^{\circ}$, respectively.
This corresponds to a maximum potential difference of 250\,nK corresponding to a gravitational shift of only $\Delta z = 2.4\,\mu$m across the ring.

We also studied the propagation of ultra-cold atoms in a ring without azimuthal confinement.
In order to investigate this, we accelerated nearly pure BECs of  $10^5$ atoms to an angular velocity  of $2 \pi\,10$\,rad/s. 
We then slowly removed the azimuthal trapping potential by ramping $\delta$ to zero and allowed the atomic cloud to fill the entire ring.
The atomic density distribution after 2\,s hold time and 6.3\,ms of time-of-flight expansion can be seen in Fig.\,\ref{fig:flatrings}d.
Again, we do not observe any additional heating or atom loss compared to the static trap nor do we observe any decay of the angular momentum of the atoms  over a timescale of 10\,s. 
We  use the fit based on Eq.\,\ref{equ:azimuthalFit} to analyse the atom distribution in Fig.\,\ref{fig:flatrings}d and determine the effective azimuthal modulation of the waveguide potential and find $h_1=0.003$,  $h_2=0.002$,  and a  temperature of 28\,nK. 
In a static trap this would correspond to a total variation of the waveguide potential by 189\,pK
or to a  difference in height of only 1.8\,nm over a distance of 1\,mm, i.e.\,a gravitational potential due to an angular misalignment of $2\,\mu$rad.
The reason for the difference between the moving and static atoms can be understood by studying the moving atoms  in their co-moving frame. 
The moving atoms experience a periodic modulation of the trapping potential at the rotation frequency (10\,Hz).
This perturbation is not resonant with the trapping frequency of $\omega_{\phi}=2\pi\,7.8$\,Hz, and thus leads only to micro-motion in the moving frame and a small density modulation in the rest frame.

\vspace{3mm}

In conclusion, the TAAP waveguides demonstrated here fulfil a long standing goal of atomtronics: the coherent transport of atoms and BECs over long distances and do so in non-trivial geometries. 
We have demonstrated virtually excitationless acceleration of ultra-cold thermal clouds and BECs to hypersonic velocities (Mach 16) and extremely high angular momentum ($17\,000\,\hbar$ per atom). 
The atoms were transported in an ultra-smooth atomtronic waveguide over macroscopic distances with BECs reaching distances of 15\,cm and ultra-cold thermal clouds up to 40\,cm without any sign of additional heating when compared to the stationary case.
The matterwave guides show no sign of roughness with an effective flatness smaller than our measurement sensitivity of 189\,pK or 1.8\,nm equivalent in gravitational height.
Such extremely flat effective potentials open very interesting possibilities in the study of 1D physics or ultra-low energy interactions. 
These neutral-atom ring accelerators open the way to many scientific and applied measurements: precise excitation of Landau levels in Quantum Hall states, hypersonic transport phenomena, and highly controlled collision experiments \cite{Amico2017NJOP, Roncaglia2011SR}. 
Our newly found ability to control barriers with pico-Kelvin precision will enable new regimes of conduction and tunnelling of ultra-cold atoms through mesoscopic barriers and channels \cite{Brantut2012Science,Krinner2017JoP}. 
This demonstration of ultra-smooth TAAP waveguides is an important step towards guided matterwave interferometry, which will lead to increased sensitivity  for fundamental physics and for applications such as inertial navigation and gravitational sensing using highly compact atomtronic devices.

\subsection*{Acknowledgements}

This work is supported by the project \emph{HELLAS-CH} (MIS 5002735), which is implemented under the \emph{Action for Strengthening Research and Innovation Infrastructures}, funded by the Operational Programme \emph{Competitiveness, Entrepreneurship and Innovation} (NSRF 2014-2020) and co-financed by Greece and the European Union (European Regional Development Fund). 
GV received funding from the European Union’s Horizon 2020 research and innovation programme under the Marie Skłodowska-Curie Grant Agreement No  750017.
SP and GD acknowledge financial support from the Hellenic Foundation for Research and Innovation (HFRI) and the General Secretariat and Technology (GSRT), under the HFRI PhD Fellowship grants (4823, 4794).

\subsection*{Contributions}

WK conceived the experimental ideas.
SP, HM and WK carried out the experiments, data analysis and theoretical work.
SP, HM, KP, VB, GD and WK contributed to building the experiment. 
All authors contributed to the result discussion and manuscript writing.

\subsection*{Competing interests}
The authors declare no competing financial interests.

\subsection*{Corresponding author}
Correspondence to Wolf von Klitzing (email: wvk@iesl.forth.gr).

\bibliographystyle{plain}

\newpage

\section*{Figure captions}
\begin{enumerate}

\item Figure 1. 
\textbf{Absorption images of ultra-cold thermal clouds and BECs in ring-shaped matterwave guides}. The upper row of images show the column density of the atoms in the ring-shaped waveguides. In a) the atoms are at rest, while b), c), and d) show atoms travelling at maximum velocity. 
 b) shows the atoms  moving in the acceleration potential along the waveguide. c) shows the atoms just after being released into the waveguide, and d) shows the atoms after travelling freely for 2\,s in the waveguide.
 \\The lower row shows the experimental background densities, calculated by subtracting from upper images  a smooth theoretical model of the atomic densities. 
 The complete absence of any signature of the ring in e) and h) clearly demonstrates that there is no detectable roughness in the atomic distribution and this in wave-guide potential. 
 From the fits to a) and d) we deduce a maximum effective modulation of the potential of 250\,nK and 189\,pK respectively.

\item Figure 2. 
\textbf{Accelerating BECs}: BEC angular position vs time in a tilted ring trap for five different angular accelerations. $x, y$ modulation amplitude is 0.55 G and 1.4 G along the $z$ axis. The lines are theoretical prediction of the position of the clouds $(\phi=\frac{1}{2}\ddot\phi t^2)$. 
 The error in the data is smaller than the point size. 
 The uncertainty in the fitted angular position and distance travelled result from an uncertainty in the position of the center of the ring and the magnification of the imaging system  (27\,mrad or 2\% of distance).
The continuation of the solid line and its corresponding data points is plotted in Fig.\,\ref{fig:transport}.

\item Figure 3. 
\textbf{Long distance transport in the accelerator ring.} a) The angular position of the condensate and thermal cloud during  14.3\,s of transport in the matterwave guide. The red line depicts the programmed trajectory of $2\pi\,10$\,rad/s. Inset shows the bi-modal distribution of the Bose-Einstein condensate after 4.1\,s of transport and 24\,ms time-of-flight expansion. b) is the condensate angular position relative to the programmed trajectory at $2\pi\,10\,\text{rad/s}$. The oscillations are partially due to a small azimuthal modulation of the trapping potential and partially due to a small center of mass oscillation of the cloud relative to the moving trap.
\end{enumerate}
\newpage

\section*{Methods}

\subsection*{Preparation of the initial atom cloud}

Following reference\,\,\cite{LIN2009PRA}, we first load $^{87}$Rb atoms in the $|F=1,\,m_\text{F}=-1\!>$ state from a magneto-optical trap into a magnetic quadrupole trap. After performing radio frequency evaporation, we transfer the atom cloud into a hybrid trap consisting of a weak quadrupole field and a crossed-beam optical dipole trap.
We evaporatively cool the atoms down to  50\,nK resulting in BECs of $3\times10^5$ atoms which we then adiabatically transfer into a tilted ring trap simply by ramping down the power of the dipole beams. 
We do not observe any significant heating, atom number loss, shape or centre of mass oscillation due to the transfer sequence.
The parameters of the final ring trap are 
$\alpha=70$\,G/cm,  
$B_{\text{m}}=1.4$\,G,  
$\delta=0.37$, 
$\omega_{\text{m}}/2\pi=5.02$\,kHz, and
$\omega_{\text{rf}}/2\pi=2.55$\,MHz.
The measured radial, axial and the azimuthal trapping frequencies of the tilted ring trap are 85.3(4)\,Hz and 46.2(3)\,Hz, and 9.17(3)\,Hz, respectively.
From the radial trapping frequency we calculate the rf-coupling strength 
$\Omega_{\text{rf}}
\equiv |g_{F}| \mu_\text{B} B_{\text{rf}}/\hbar
\approx 2\pi\,357$\,kHz.

\subsection*{Absorption imaging and fitting}

The images of Fig.\,\ref{fig:flatrings} were taken by absorption imaging, where one sends resonant light through the atom cloud. 
The transmitted light is then imaged onto a CCD camera, from which one can then deduce the column density of the atoms via a modified version of the Beer–Lambert law \cite{Pappa2011NJOP}.
In Fig.\,\ref{fig:flatrings}a) the atoms are nearly at rest. 
For the right three image columns in Fig.\,\ref{fig:flatrings}, i.e.~figures b), c) and d) we accelerated the atoms to an angular speed of  $2\pi 10$\,rad/s in the clockwise direction and guided the atoms for a duration of 0.2\,s, 0.452\,s and 2\,s, respectively. 
The difference in temperature between the static and travelling atom clouds (430\,nK and 28\,nK, respectively) was chosen high enough such that the clouds fill the entire ring and yet    low enough for the modulation in density due to the potential differences to be large.
For the fit we use a harmonic potential offset by the radius of the ring, and azimuthally the potential described in eq.\,\ref{equ:azimuthalFit}:
\[
U_\rho=\left(\frac{\rho-\rho_0}{\Delta \rho}\right)^2,
\]\[
U_\phi=1+h_1 \cos (\phi +\phi_1)+h_2 \cos (2 \phi +\phi_2),
\]
where $\rho$ and $\phi$ are the spatial coordinates of the image and $\rho_0$ the radius of the ring. 
\[
U=U_\rho + U_\phi
\]
The fitted optical density is then a bimodal distribution using the above potential:
\[
od=j_0 \exp[-U/T] + k_0 \text{Re}[(1 - U/\mu)^{3/2}]\,,
\]
where $j_0$ ($k_0$) is the  peak radial  optical density of the thermal (condensate) cloud averaged azimuthally along the ring.
We also allowed for a fit of the ellipticity of the ring, but found it to be negligible ($\leq 10^{-3}$).

The fit parameters have arbitrary scales, which were chosen for numerical convenience and are taken into account in the calculation of the physical parameters.
The column densities of the atoms follow from the optical densities their absorption properties including saturation \cite{Pappa2011NJOP}.
We determine the absolute energy scale $T$ of the fit  from an independent measurement of the radial trapping frequency and/or from the temperature of the atoms using  time-of-flight expansion of the cloud.

\subsection*{Flatness of the ring potential}

The strong suppression of the potential variations present in the flat ring finds its origin in the fact that the kinetic energy scales with the square of the velocity whereas the density scales linearly with the velocity. We derived an equation predicting the azimuthal density variations due to these barriers with a rotating atomic cloud. A simple treatment is to assume a flow with speed $v$ experiencing a dip or hill in the trap, whilst keeping the flux constant along the waveguide. The flux is the product of flow velocity $v$ and the density $n$ and it is constant in the case of a steady-state flow along the waveguide. The kinetic energy decreases (increases) at the top (bottom) of the hill exactly by the difference in the potential energy. The fractional change in $v$ and equivalently in the case of steady flow the fractional change in atomic density due to a barrier of height $\Delta$E, at an angular speed $\Omega$ in a static potential of radius $R$, is $\Delta v/v = \Delta n/n = -1+\sqrt[]{1+\frac{2 \Delta E}{mv^{2}}}$ where $v=R\Omega$. For a $2 \pi \times 10$ rad/s angular speed the calculated variation in the density is 0.001\%, and 0.0006\% for the $\phi$ and 2$\phi$ tilt (see \ref{fig:flatrings}d), respectively. We find that the angular position of the minimum (maximum) density is randomly distributed over the entire 2$\pi$ shot-to-shot. The measured sample temperature is 28\,nK and thus the effective flatness of the atomic density in the waveguide is $\sim$ 189\,pK. Using $k_{\text{B}}T = mgh$, 189\,pK is equivalent to 1.8\,nm height difference in the gravitational field along the waveguide.
This is a strong indication that the density variation we observe is not due to the static ring potential but a limit of our detection.

\subsection*{Acceleration of the atoms}

Optimal control theory provides a simple solution to the smooth acceleration of atoms in harmonic traps, the so-called 
\emph{`bang-bang'} scheme. This works by compensating  the force due to a constant acceleration by an opposite force due to an offset in the position of the atomic cloud relative to the center of the moving harmonic trap. 
Concretely, one would jump the position of the trap forward, right before acceleration starts. This would result in a sudden angular displacement of the harmonic potential.
In the accelerated frame, the atoms then stay  exactly at the bottom of the accelerated, effective trapping potential.
To calculate the jump amplitude, we equate the force due to the trap acceleration $\ddot\phi$ and gradient of the potential at a distance $\Delta \phi$ away from the trap center, $m \omega{^2} \Delta \phi = m \ddot\phi$, where m is the mass of the particle and $\omega$ the trapping frequency. The amplitude of the trap jump is then, $\Delta \phi = + \ddot\phi//\omega{^2}$.

Care has to be taken that the atomic cloud remains in the harmonic region of the trap.  For very elongated clouds in the absence of collisions (like the ones in the case at hand), this condition can be relaxed somewhat to require only that the trap be harmonic over the size of the cloud and that the rate of change of the trapping frequencies be slow ($\dot\omega<\omega^2$).  This ensures that no shape-oscillations are excited and that the cloud remains self-similar (Gaussian for the thermal cloud and parabolic for the condensate). The maximum attainable angular frequency for the confinement of the atomic sample in the ring is limited by the radial quadrupole gradient, which sets an upper limit for the centripetal force. At very high acceleration this limit can be easily breached.

\subsection*{Oscillations in the moving trap}

The angular position of the atomic cloud in Fig.\,\ref{fig:transport}b) was fitted for the entire data set to
\[
\phi(t)=\phi_0 + 2\pi 10\,t + a_1 \sin(2\pi10\,t+\phi_{1})+ a_2 \sin(2\pi20\,t+\phi_{2})+ a_3 e^{-t/\tau} \sin(2\pi \omega_\phi\,t+\phi_{3}).
\]
This corresponds to a phase offset $(\phi_0)$ of an ideal motion $(2\pi 10\,t )$ and a micro-motion at the same frequency as the forced rotation around the ring $( a_1 \sin(2\pi10\,t)$ and at its first harmonic $( a_2 \sin(2\pi20\,t)$.
We find the oscillation amplitudes $a_1$, $a_2$ and $a_3$ to be 140 (10), 40 (10) and 70 (10)\,mrad, respectively.
The fit yields $\phi_0$, $\phi_1$, $\phi_2$ and $\phi_3$ equal to 2.2, 3.3, 3.5 and 0.5\,rad, respectively. 
No drift in phase or amplitude within our estimated resolution was detected for the ten and twenty Hertz oscillations over the full 14.3\,s.
The final term $(a_3 e^{-t/\tau} \sin[2\pi \omega_\phi\,t+\phi_{3}])$ is a centre of mass damped oscillation of the atomic cloud in the moving trap.
The decay time constant $(\tau)$ for the azimuthal trap oscillation at 7.76 (1)\,Hz in the moving trap is 5.3\,s.

\subsection*{Internal coherence of the BEC}

A zero-temperature BEC has a flat phase distribution at rest and a simple gradient when in motion. 
At higher temperatures, however, thermal excitations can occur, which then result in random phase gradients \cite{Petrov2001PRL}.
Phase gradients correspond to a flow of the condensate.
Inside the trap, the superfluidity strongly suppresses  density fluctuations from being generated by these phase fluctuations.
In time-of-flight free expansion, however, the density of the condensates drops very quickly and with it the critical superfluid velocity.
Once the critical velocity drops to below the velocity associated with the phase fluctuations density patterns (usually stripes for elongated BECs) form.
Such density patterns have been observed experimentally \cite{Dettmer2001PRL}.
We do not observe any such fringes up to our maximum expansion time of 24\,ms.

\subsection*{Speed of sound in the BEC}

The motion of a BEC in a rough potential has two completely different regimes: At very low velocities, the superfluid motion prevails, where the BEC flows around any obstacle or roughness without picking up any excitation. 
At higher velocities, a transport along potentials with higher spatial frequencies causes phonon-like excitations in the condensate. 
The border between the two regimes is the Bogoliubov speed of sound \cite{kavoulakis1998}:

\begin{equation}
c=\sqrt{\frac{n \,U_{0}}{M}}\,,
\label{eq:speedofsound}
\end{equation}
where $U_{0}=4 \pi \hbar^{2} a_\text{Rb}/M$ is the bosonic interaction parameter, $a_\text{Rb}$ and M are the scattering length and  mass of $^{87}$Rb, respectively. The density of atoms in the BEC is  $n$.
In our case, the density of the gas takes the parabolic shape of the trap. The maximum speed of sound occurs at the center of the trap, where the peak density $n=n_\text{max}$ is 
\begin{equation}
n_\text{max}=\frac{1}{8\pi}\left[15 N \left(\frac{M \omega_\text{ho}}{\hbar \sqrt{a_\text{Rb}}}\right)^{3}\right]^{2/5}\,\,,
\end{equation}
where  $\omega_\text{ho}=\left(\omega_x  \omega_y  \omega_z\right)^{1/3}=2\pi\left(46\,\text{Hz}\times 85\,\text{Hz}\times 7.8\,\text{Hz}\right)^{1/3}=31.2\,\text{Hz}$. 
For $N=3\times10^{5}$ atoms this results in a maximum speed of sound in the BEC of  $c_\text{max}=1.75$\,mm/s.

The condensate travels along the ring at a speed of $v= \omega\times R =2 \pi 10\,\text{Hz}\times 443\mu\text{m}=27.8$\,mm/s.
The condensates therefore travel at a speed of 16 times their peak speed of sound (Mach number o $M_\text{min}=v/c_\text{max}= 16$), well above the velocity at which their superfluidity allows them to flow frictionless around defects in the waveguide. 
The absence of any heating associated with their motion thus experimentally proofs the extreme smoothness of the TAAP waveguides.

\subsection*{Angular momentum of the guided atoms}

In the case of moving atoms in a flat ring waveguide (Fig.\,\ref{fig:flatrings}d), we measure the angular momentum via the time-of-flight method. The ring potential is suddenly switched off and we let the cloud fall under gravity for different time-of-flights. Due to the in-trap angular momentum the cloud radius increases with the time-of-flight. We fit the cloud radius to $R(t) = R(0) \sqrt{1+(\Omega t)^{2}}$ where $R(0)$ is the in-trap radius at an angular speed of $\Omega$.  After 1\,s of hold time in the flat ring waveguide, we measure an angular speed of $2 \pi 10.01 (\pm 0.06)$\,rad/s while the programmed $\Omega$ is $2 \pi 10$\,rad/s.

\subsection*{Radius of an annular atom cloud under rotation}

The centrifugal force on the atoms together with their harmonic radial confinement cause the radius of the waveguide to increase.
The  radius of the ring increases as a function of the angular speed according to $R (\Omega) = R_0 (1-\Omega^2/\omega^2_{\text{r}})^{-1}$, where $\omega_{\text{r}}$ is the radial trapping frequency, $\Omega$ is the angular speed and $R_0$ is the ring radius at $\Omega=0$.
In line with this prediction, the ring radius increases from $436(2)\mu$m in a static ring to $443.4(4) \mu$m at an angular speed of $2\pi\times 10$\,rad/s.

% HM 190214: bibliography style from wlscirep.cls

\end{document}